\documentclass[amsmath,amssymb]{revtex4}

\usepackage{graphicx}
\usepackage{dcolumn}
\usepackage{bm}

\begin{document}

\title{$PT$-Symmetric Matrix Quantum Mechanics}

\author{Peter N. Meisinger}
\author{Michael C. Ogilvie}%
 \email{mco@wuphys.wustl.edu}
\affiliation{%
Department of Physics, Washington University, St.\ Louis, MO 63130, USA
}%

\date{\today}

\begin{abstract}
Recently developed methods for PT-symmetric models are applied to
quantum-mechanical matrix models. We consider in detail the case of
potentials of the form $V=-(g/N^{p/2-1})Tr\left(iM\right)^{p}$ and
show how the calculation of all singlet wave functions can be reduced
to solving a one-dimensional PT-symmetric model. The large-$N$ limit
of this class of models exists, and properties of the lowest-lying
singlet state can be computed using WKB. For $p=3,4$, the energy
of this state for small values of $N$ appears to show rapid convergence
to the large-$N$ limit. For the special case of $p=4$, we extend
recent work on the $-gx^{4}$ potential to the matrix model: we show
that the PT-symmetric matrix model is equivalent to a hermitian matrix
model with a potential proportional to $+(4g/N)Tr\Pi^{4}$. However,
this hermitian equivalent model includes an anomaly term $\hbar\sqrt{2g/N}Tr\Pi$.
In the large-$N$ limit, the anomaly term does not contribute at leading
order to the properties of singlet states. 
\end{abstract}

\maketitle

\section{Introduction}

Matrix models appear in many contexts in modern theoretical physics,
with applications ranging from condensed matter physics to string
theory. Interest in the large-$N$ limit of matrix models was strongly
motivated by work on the large-$N_{c}$ limit of QCD  \cite{'tHooft:1973jz},
but interest today is much wider. For example, Hermitian
matrix quantum mechanics leads to a construction of two-dimensional quantum gravity
coupled to $c=1$ matter \cite{Kazakov:1988ch}.

We will show below that the matrix techniques pioneered in 
\cite{Brezin:1977sv} for Hermitian matrix quantum mechanics
can be extended to $PT$-symmetric matrix quantum mechanics,
where the matrices are normal but not necessarily Hermitian.
The large-$N$ limit can then be taken in $PT$-symmetric matrix theories
just as in the Hermitian case. Quantities of interest such as the
scaled ground state energy and scaled moments can be calculated using
WKB methods. In the special case of a quartic potential with the {}``wrong''
sign, we prove using functional integration for all values of $N$
 that the $PT$-symmetric
model is equivalent to a hermitian matrix model with an anomaly, as
in the one-component case \cite{Bender:2006wt,Jones:2006et}. Interestingly,
the anomaly vanishes to leading order in the large-$N$ limit.

\section{Formalism}

The solution for all $N$ of the quantum mechanics problem associated with the
Euclidean Lagrangian\begin{equation}
L=\frac{1}{2}Tr\left(\frac{dM}{dt}\right)^{2}+\frac{g}{N}TrM^{4}\end{equation}
where $M$ is an $N\times N$ Hermitian matrix was first given by
Brezin et al.\cite{Brezin:1977sv}.
The ground state $\psi$
is a symmetric function of the eigenvalues $\lambda_{j}$ of $M$.
The antisymmetric wave function $\phi$ defined by\begin{equation}
\phi\left(\lambda_{1},..,\lambda_{N}\right)=\left[\prod_{j<k}\left(\lambda_{j}-\lambda_{k}\right)\right]\psi\left(\lambda_{1},..,\lambda_{N}\right)\end{equation}
satisfies the Schrodinger equation\begin{equation}
\sum_{j}\left[-\frac{1}{2}\frac{\partial^{2}}{\partial\lambda_{j}^{2}}+\frac{g}{N}\lambda_{j}^{4}\right]\phi=N^{2}E^{(0)}\phi\end{equation}
where $E^{(0)}$is the ground state energy scaled for the large-$N$
limit. This equation separates into $N$ individual Schrodinger equations,
one for each eigenvalue, and the antisymmetry of $\phi$ determines
$N^{2}E^{(0)}$ as the sum of the $N$ lowest eigenvalues.

Here we solve the corresponding problem where the potential term is
$PT$-symmetric but not Hermitian. As shown by Bender and
Boettcher \cite{Bender:1998ke}, the
one-variable problem may be solved by extending the coordinate variable
into the complex plane. This implies that for $PT$-symmetric matrix
problems, we must analytically continue the eigenvalues of $M$ into
the complex plane, and in general $M$ will be normal rather than
Hermitian. We consider the Euclidean Lagrangian

\begin{equation}
L=\frac{1}{2}Tr\left(\frac{dM}{dt}\right)^{2}-\frac{g}{N^{p/2-1}}Tr\left(iM\right)^{p}\end{equation}
with $g>0$. Making the substitution $M\rightarrow U\Lambda U^{+}$,
with $U$ unitary and $\Lambda$ diagonal, we can write $L$ as\begin{equation}
L=\frac{1}{2}\sum_{j}\left(\frac{d\lambda_{j}}{dt}\right)^{2}+\sum_{j,k}\frac{1}{2}\left(\lambda_{j}-\lambda_{k}\right)^{2}\left(\frac{dH}{dt}\right)_{jk}\left(\frac{dH}{dt}\right)_{kj}-\frac{g}{N^{p/2-1}}\sum_{j}\left(i\lambda_{j}\right)^{p}
\end{equation}
where\begin{equation}
\frac{dH}{dt}=-iU^{+}\frac{dU}{dt}.\end{equation}
In the analysis of conventional matrix models by Brezin et al., a
variational argument shows that the ground state is a singlet, with
no dependence on $U$. Because the $\lambda_{j}$'s are
in general complex
for $PT$-symmetric theories, this
argument does not apply. However, in two cases we can prove
that the ground state is indeed a singlet: for $p=2$, which is trivial,
and for $p=4$, where the explicit equivalence with a hermitian matrix
model proven below can be used. Henceforth, we will assume that the
ground state is a singlet, but our results will apply in any case
to the lowest-energy singlet state.

We have now reduced the problem of finding the ground state to the
problem of solving for the first $N$ states of the single-variable Hamiltonian
\begin{equation}
H=\frac{1}{2}p^{2}-\frac{g}{N^{p/2-1}}\left(i\lambda\right)^{p}.\end{equation}
This Hamiltonian is $PT$-symmetric but in general not Hermitian.
The case $p=2$ is the simple harmonic oscillator. 
For $p>2$, the Schrodinger equation associated with each eigenvalue may
be continued into the complex plane as explained
in \cite{Bender:1998ke}.
We exclude the case $p<2$, where $PT$ symmetry is spontaneously broken
and the eigenvalues of $H$ are no longer real.

\section{Ground State Properties}

As with Hermitian matrix models. the ground state energy is the sum
of the first $N$ eigenenergies of the Hamiltonian $H$. In the
large $N$ limit, this sum may be calculated using WKB.
A novelty of WKB for $PT$-symmetric models is the extension
of classical paths into the complex plane.
This topic has been treated extensively  in 
\cite{Bender:1998ke,Bender:1998gh}.

We define the Fermi energy $E_{F}$ as the energy of the $N$'th state
\begin{eqnarray}
N & = & \frac{1}{2\pi}\int dpd\lambda\,\theta\left[E_{F}-H(p,\lambda)\right]\end{eqnarray}
where the path of integration must be a closed, classical path in the
complex $p-\lambda$ plane. In order to construct the large-$N$ limit,
we perform the rescaling $p\rightarrow\sqrt{N}p$ and $\lambda\rightarrow\sqrt{N}\lambda$
yielding
\begin{equation}
H_{sc}(p,\lambda)=\frac{1}{2}p^{2}-g\left(i\lambda\right)^{p}\end{equation}
where the scaled Hamiltonian $H_{sc}$ is related to $H$ by  $H=NH_{sc}$.
We introduce a rescaled Fermi energy
 $\epsilon_{F}$ given by $E_{F}=N\epsilon_{F}$, which is implicitly
defined by 
\begin{equation}
1=\frac{1}{2\pi}\int dpd\lambda\theta\left[\epsilon_{F}-H_{sc}(p,\lambda)\right].
\label{eqn:one}\end{equation}
After carrying out the integration over $p$, we have\begin{equation}
1=\frac{1}{\pi}\int d\lambda\sqrt{2\epsilon_{F}+2g\left(i\lambda\right)^{p}}\theta\left[\epsilon_{F}+g\left(i\lambda\right)^{p}\right]\end{equation}
where the contour of integration is taken along a path between the
turning points which are the analytic continuation of the turning
points at $p=2$. This equation determines $\epsilon_{F}$ as a function
of $g$. 

We define a scaled ground state energy $E^{(0)}$ by
\begin{equation}
E^{(0)}_N=\frac{1}{N^{2}}\sum_{k=0}^{N-1}E_{k}.\end{equation}
The WKB result for the sum of the energies less than $E_{F}$ can be written
as \begin{equation}
\sum_{k=0}^{N-1}E_{k}=\frac{N^{2}}{2\pi}\int dpd\lambda\, H_{sc}(p,\lambda)\theta\left[\epsilon_{F}-H_{sc}(p,\lambda)\right]\end{equation}
so that in the large-$N$ limit $E^{(0)}_{\infty}$ is given by
\begin{equation}
E^{(0)}_{\infty}=
\frac{1}{2\pi}\int dpd\lambda\,H_{sc}(p,\lambda)\theta\left[\epsilon_{F}-H_{sc}(p,\lambda)\right]\end{equation}
The integration over $p$ is facilitated by using  equation (\ref{eqn:one}) to insert
a factor of $\epsilon_{F}$, giving
\begin{equation}
E^{(0)}_{\infty}=
\epsilon_{F}-\frac{1}{2\pi}\int dpd\lambda\left[\epsilon_{F}-H_{sc}(p,\lambda)\right]\theta\left[\epsilon_{F}-H_{sc}(p,\lambda)\right].\end{equation}
The
integral over $p$ then yields
\begin{equation}
E^{(0)}_{\infty}=\epsilon_{F}-\frac{1}{3\pi}\int d\lambda\left[2\epsilon_{F}+2g\left(i\lambda\right)^{p}\right]^{3/2}\theta\left[\epsilon_{F}+g\left(i\lambda\right)^{p}\right].
\label{eqn:E0}\end{equation}

The turning points in the complex $\lambda$ plane are\begin{equation}
\lambda_{-}=\left(\frac{\epsilon_{F}}{g}\right)^{1/p}e^{i\pi\left(3/2-1/p\right)}\end{equation}
\begin{equation}
\lambda_{+}=\left(\frac{\epsilon_{F}}{g}\right)^{1/p}e^{-i\pi\left(1/2-1/p\right)}\end{equation}
We integrate $\lambda$ along a two-segment, straight-line path
connecting the two turning
points via the origin \cite{Bender:1998ke}. 
Solving equation  (\ref{eqn:one})   for $\epsilon_{F}$, we
find\begin{equation}
\epsilon_{F}=
\left[\left(\frac{\pi}{2}\right)^{p}\left(\frac{\Gamma(3/2+1/p)}{\sin\left(\pi/p\right)\Gamma(1+1/p)}\right)^{2p}g^{2}\right]^{\frac{1}{p+2}},
\end{equation}
and solving (\ref{eqn:E0}) for the scaled ground state energy we have\begin{equation}
E^{(0)}_\infty=\frac{p+2}{3p+2}\epsilon_{F}
=\frac{p+2}{3p+2}\left[\left(\frac{\pi}{2}\right)^{p}\left(\frac{\Gamma(3/2+1/p)}{\sin\left(\pi/p\right)\Gamma(1+1/p)}\right)^{2p}g^{2}\right]^{\frac{1}{p+2}}.\end{equation}
 For $p=2$, this evaluates to $E^{(0)}=\sqrt{g/2}$ , in agreement
 with the explicit result for the harmonic oscillator.
 
It is very interesting to compare the large-$N$ result with results
for finite $N$. The low-lying eigenvalues for the Hamiltonian $p^{2}-(ix)^{p}$
have been calculated by Bender and Boettcher in \cite{Bender:1998ke} for the cases $p=3$ and
$p=4$; the case $p=2$ is trivial. We can use their results by noting
that the eigenvalues of our Hamiltonian $H$
 are related to theirs
by\begin{equation}
E_{j}=\frac{g^{2/(p+2)}}{2^{p/(p+2)}N^{(p-2)/(p+2)}}E_{j}^{BB}.\end{equation}
 Results for $p=3$ and $4$ and small values of $N$ are compared
with the large-$N$ limit in Table 1. The energies for finite values of $N$ rapidly approach the $N \to \infty$ limit. The approach to the limit appears monotonic in both cases, but with opposite sign.

\hfill
\begin{table}
\begin{center}
\begin{tabular}{|c||c|c|}
\hline 
N&
p=3&
p=4\tabularnewline
\hline
1&
0.762852&
0.930546\tabularnewline
\hline 
2&
0.756058&
0.935067\tabularnewline
\hline 
3&
0.75486&
0.935846\tabularnewline
\hline 
4&
0.754443&
0.936115\tabularnewline
\hline 
5&
0.754251&
0.936239\tabularnewline
\hline 
6&
0.754147&
0.936306\tabularnewline
\hline 
7&
0.754084&
0.936347\tabularnewline
\hline
8&
0.754043&
0.936372\tabularnewline
\hline
$\infty$&
0.753991&
0.936458\tabularnewline
\hline
\end{tabular}
\end{center}
\caption{The scaled ground state energy $E^{(0)}_N$ at $g=1$ for $p=3$ and $p=4$.}
\end{table}

The expected value of $\left\langle TrM\right\rangle $ for large $N$
is given by\begin{eqnarray}
\left\langle TrM\right\rangle =\sum_{j=0}^{N-1}\left\langle \lambda_{j}\right\rangle  & = & \frac{1}{2\pi}\int dpd\lambda\,\lambda\theta\left[E_{F}-H(p,\lambda)\right].\end{eqnarray}
Calculations of higher moments $\left\langle TrM^{n}\right\rangle $
are carried out in the same manner.
Upon rescaling, we find that $\left\langle TrM\right\rangle $ grows
as $N^{3/2},$ and the scaled expectation value is given by 

\begin{eqnarray}
\mu=\lim_{N\rightarrow\infty}\frac{1}{N^{3/2}}\left\langle TrM\right\rangle  & = & \frac{1}{2\pi}\int dpd\lambda\,\lambda\theta\left[\epsilon_{F}-H_{sc}(p,\lambda)\right]\end{eqnarray}
which reduces to \begin{equation}
\mu=\frac{1}{\pi}\int d\lambda\,\lambda\sqrt{2\epsilon_{F}+2g\left(i\lambda\right)^{p}}\theta\left[2\epsilon_{F}+2g\left(i\lambda\right)^{p}\right].\end{equation}
Using the same two-segment straight line path as before, we find that
 \begin{equation}
\mu=-i\left(\frac{\pi}{2g}\right)^{\frac{1}{p+2}}\sin\left(\frac{\pi}{p}\right)^{-\frac{2}{p+2}}\cos\left(\frac{\pi}{p}\right)\left[\frac{\Gamma(3/2+1/p)}{\Gamma(1+1/p)}\right]^{\frac{p+4}{p+2}}\frac{\Gamma\left(1+2/p\right)}{\Gamma\left(3/2+2/p\right)}.\end{equation}
For $p=2$, $\mu=0$, as expected for a harmonic oscillator. For $p>2$,
the expectation value $\mu$ is imaginary because $\left\langle \lambda_{j}\right\rangle $
for each eigenstate of the reduced problem is imaginary \cite{Bender:1998ke}.
For $p=3$, $\mu=-0.52006i$. For $p=4$, $\mu=-0.772539i$. In the
limit $p\rightarrow\infty$, $\mu$ goes to -i. This behavior is easy
to understand, because in this limit, the turning points become degenerate
at $-i$. 

\section{Special case of $TrM^{4}$}

For the case of a $TrM^{4}$ interaction, we can explicitly exhibit the equivalence
of the PT-symmetric matrix model with a conventional Hermitian quantum mechanical
system. As in the single-variable case, there is a parity-violating
anomaly, in the form of an extra term in the Hermitian form of the
Hamiltonian, proportional to $\hbar$. We show below that the anomaly
term does not contribute at leading order in the large-$N$ limit.

The derivation of the equivalence closely follows the path integral
derivation for the single-variable case \cite{Bender:2006wt,Jones:2006et}. 
The Euclidean Lagrangian is

\begin{equation}
L=\frac{1}{2}Tr\left(\frac{dM}{dt}\right)^{2}+\frac{1}{2}m^{2}Tr\, M^{2}-\frac{g}{N}TrM^{4}\end{equation}
 and the path integral expression for the partition function is\begin{equation}
Z=\int\left[dM\right]exp\left\{ -\int dt\, L\right\} .\end{equation}
Motivated by the case of a single variable, we make the substitution\begin{equation}
M=-2i\sqrt{1+iH}\end{equation}
where $H$ is an Hermitian matrix. Because $M$ and $H$ are simultaneously
diagonalizable, this transformation is tantamount to the relation\begin{equation}
\lambda_{j}=-2i\sqrt{1+ih_{j}}\end{equation}
between the eigenvalues of $M$ and the eigenvalues $h_{j}$ of $H$.
The change of variables induces a measure factor\begin{equation}
[dM]=\frac{[dH]}{Det[\sqrt{1+iH}]}\end{equation}
 where the functional determinant depends only on the eigenvalues
of $H$. The Lagrangian becomes\begin{equation}
L=\frac{1}{2}Tr\frac{(dH/dt)^{2}}{1+iH}-2m^{2}Tr\,(1+iH)-16\frac{g}{N}Tr\,(1+iH)^{2}\end{equation}
at the classical level. However, following \cite{Jones:2006et}, we note that in
the matrix case the change of variables introduces an extra term in
the potential of the form\begin{equation}
\Delta V=\sum_{j}\frac{1}{8}\left[\frac{d}{dh_{j}}\left(\frac{dh_{j}}{d\lambda_{j}}\right)\right]^{2}\end{equation}
which can be written as\begin{equation}
\Delta V=-\frac{1}{32}\sum_{j}\frac{1}{1+ih_{j}}=-\frac{1}{32}Tr\left(\frac{1}{1+iH}\right).\end{equation}
The partition function is now \begin{equation}
Z=\int\frac{\left[dH\right]}{\det\left[\sqrt{1+iH}\right]}exp\left\{ -\int dt\left[\frac{1}{2}Tr\frac{(dH/dt)^{2}}{1+iH}-2m^{2}Tr(1+iH)-\frac{16g}{N}Tr(1+iH)^{2}-\frac{1}{32}Tr\left(\frac{1}{1+iH}\right)\right]\right\} \end{equation}
 We introduce a hermitian matrix-valued field $\Pi$ using the identity\begin{equation}
\frac{1}{\det\left[\sqrt{1+iH}\right]}=\int\left[d\Pi\right]exp\left\{ -\int dt\, Tr\left[\frac{1}{2}\left(1+iH\right)\left(\Pi-\frac{\dot{iH}+1/4}{1+iH}\right)^{2}\right]\right\} .\end{equation}
Dropping and adding appropriate total derivatives and integrating
by parts yields 
\begin{equation}
Z=\int\left[dH\right]\left[d\Pi\right]exp\left\{ -\int dt\, Tr\left[-2m^{2}(1+iH)-16\frac{g}{N}(1+iH)^{2}+\frac{1}{2}\left(1+iH\right)\Pi^{2}+\dot{\Pi}(1+iH)-\frac{1}{4}\Pi\right]\right\} \end{equation}
The integration over $H$ is Gaussian, and the shift $H\rightarrow H+i$
gives
\begin{equation}
Z=\int\left[d\Pi\right]exp\left\{ -\int dt\, Tr\left[\frac{N}{64g}\left(\dot{\Pi}^{2}-2m^{2}\Pi^{2}+\frac{1}{4}\Pi^{4}\right)-\frac{1}{4}\Pi\right]\right\} .\end{equation}
After the rescaling $\Pi\rightarrow\sqrt{32g/N}\Pi$ we have finally\begin{equation}
Z=\int\left[d\Pi\right]exp\left\{ -\int dt\, Tr\left[\frac{1}{2}\left(\dot{\Pi}^{2}-2m^{2}\Pi^{2}\right)+\frac{4g}{N}\Pi^{4}-\sqrt{2g/N}\Pi\right]\right\} \end{equation}

We have now proven the equivalence of the PT-symmetric matrix model
defined by\begin{equation}
L=\frac{1}{2}Tr\left(\frac{dM}{dt}\right)^{2}+\frac{1}{2}m^{2}Tr\, M^{2}-\frac{g}{N}TrM^{4}\end{equation}
to the conventional quantum mechanics matrix model given by\begin{equation}
L'=\frac{1}{2}Tr\left(\frac{d\Pi}{dt}\right)^{2}-\sqrt{\frac{2g}{N}}Tr\Pi-m^{2}Tr\Pi^{2}+\frac{4g}{N}Tr\Pi^{4}.\end{equation}
 This equivalence implies that the energy eigenvalues of the corresponding
 Hamiltonians are the same. 
This could also be proven using the single-variable
equivalence for the special case of singlet states, but the functional
integral proof encompasses both singlet and non-singlet states at
once. The equivalence of these two models also allows for an easy
proof of the singlet nature of the ground state.
Standard variational arguments show that the ground state of
the Hermitian form is a singlet. The direct quantum mechanical equivalence
of the single-variable case is then sufficient to prove that the ground
state of the $PT$-symmetric form is also a singlet.

As in the single-variable case, there is a linear term
of order $\hbar$ appearing in the Lagrangian and Hamiltonian of the
Hermitian form of the model. This term represents
a quantum mechanical anomaly special
to the $TrM^{4}$ model. 
To determine the fate of the anomaly in the large-$N$ limit, 
we construct the scaled Hamiltonian 
of the Hermitian form in exactly the same way as for
the $PT$-symmetric
form. It is given by \begin{equation}
H_{sc}=\frac{1}{2}p^{2}-\frac{1}{N}\sqrt{2g}x-m^{2}x^{2}+4gx^{4},\end{equation}
indicating that the effect of the anomaly is absent in leading order
of the large-$N$ expansion. One easily checks for the $m=0$ case
that the Hermitian form without the linear term reproduces
the $PT$-symmetric prediction for $E^{(0)}_\infty$
at $p=4$.

\begin{acknowledgments}
The authors would like to thank Carl M. Bender for useful discussions,
and Stefan Boettcher for providing data used in constructing Table 1.
We gratefully acknowledge the support of the U.S.\ Department of Energy.
\end{acknowledgments}

\end{document}